\documentclass{WileyMSP-template}
\begin{document}

\pagestyle{fancy}

\title{Enhancement of Circular Dichroism in Chiral Dielectric Metasurfaces by Ion Beam Irradiation }

\maketitle


\author{Anna Fitriana,*}
\author{Katsuya Tanaka,}
\author{Lukas Raam Jaeger,}
\author{Martin Hafermann,}
\author{Thomas Pertsch,}
\author{Carsten Ronning,}
\author{Isabelle Staude}




\begin{affiliations}

A. Fitriana, K. Tanaka, L. Raam Jaeger, M. Hafermann, C. Ronning, I. Staude\\
Institute of Solid State Physics\\
Friedrich-Schiller-University Jena\\
Max-Wien-Platz 1, 07743 Jena, Germany\\
Email: anna.fitriana@uni-jena.de\\
\vspace{5mm} 
A. Fitriana, K. Tanaka, T. Pertsch, C. Ronning, I. Staude\\
Abbe Center of Photonics\\
Friedrich-Schiller-University Jena\\
Albert-Einstein-Straße 6, 07745 Jena, Germany\\
\vspace{5mm} 
A. Fitriana, L.R. Jaeger\\
ARC Centre of Excellence for Transformative Meta-Optical Systems (TMOS)\\
Department of Electronic Materials Engineering, Research School of Physics\\
Australian National University\\
Canberra, 2600 ACT, Australia\\
\vspace{5mm} 
K. Tanaka, T. Pertsch, I. Staude\\
Max Planck School of Photonics\\
Hans-Knöll-Straße 1, 07745 Jena, Germany\\
\vspace{5mm} 
T. Pertsch\\
Institute of Applied Physics Physics\\
Friedrich-Schiller-University Jena\\
Albert-Einstein-Straße 6, 07745 Jena, Germany\\
\vspace{5mm} 
T. Pertsch\\
Fraunhofer Institute for Applied Optics and Precision Engineering IOF\\
Albert-Einstein-Straße 7, 07745 Jena, Germany
\end{affiliations}


\begin{abstract}
Resonant chiral dielectric metasurfaces can support circular dichroism exceeding that of natural materials, but their small dissipative losses simultaneously limit the maximization of circular dichroism, which inherently relies on absorption. Importantly, while the condition for optimal circular dichroism in resonant structures can be rigorously formulated based on the concept of critical coupling, controlling the amount of absorption experimentally, and ideally tuning it to the optimal value post-fabrication, remains elusive. Here, we experimentally tailor the dissipative losses of chiral bilayer dielectric metasurfaces post-fabrication using energetic ion beam irradiation. Specifically, we study the transmission characteristics of $C$\textsubscript{4}-symmetric chiral metasurface consisting of silicon nanocuboid arrays embedded in silica glass using polarization-resolved spectroscopy. We enhance the circular dichroism from 0.70 in the pristine, unirradiated metasurface to 0.85 after irradiation. Our experimental results are complemented by numerical simulations allowing us to retrieve the refractive index changes induced by the ion beam irradiation in the constituent materials of the metasurface. Our work offers a new approach to globally maximize optical chirality in engineered nanostructures, paving the way towards chiral emission and advanced polarization control applications.

\end{abstract}

\section{Introduction}
Circular dichroism (CD) refers to a differential absorption of left- and right-handed circularly polarized light (LCP and RCP) by chiral objects or materials. While CD naturally occurs in biological systems \cite{beychok1966circular,ranjbar2009circular} and molecular compounds \cite{tedesco2015induced,kiss2019simple}, its magnitude is typically weak. Engineered chiral metamaterials or metasurfaces \cite{decker2007circular, decker2010twisted, ma2015planar} composed of achiral materials can exhibit significantly stronger CD leveraging resonant optical response and optimized geometries. These artificial structures have been instrumental in developing a deeper understanding of chiral light-matter interactions and have important potential for advanced applications, such as ultrathin circular polarizers \cite{zhao2012twisted, zheng2022compound}, spin-selective wavefront shaping \cite{chen2018spin}, and chiral sensing \cite{wang2020controlling,ali2023dielectric}. 

Fundamentally, CD arises from asymmetric absorption of circularly polarized waves and is intrinsically linked to loss mechanisms including electronic transitions \cite{govorov2010theory, andrews2020physical}, phonon interactions \cite{nafie1976vibrational, xu2023expanding}, and other dissipative channels. Maximizing CD in resonant chiral metasurfaces for a given chiral design requires a balance between a non-radiative decay rate $\gamma_d$ associated with intrinsic loss and a radiative decay rate $\gamma_r$, a condition referred as a critical coupling  \cite{gorkunov2020metasurfaces}. 

Recent advances in nanofabrication technology, particularly state-of-the-art lithographic techniques such as electron-beam lithography and focused ion beam milling, have enabled high-resolution patterning of 2D and 3D structures on diverse material platforms \cite{hendry2010ultrasensitive,hentschel2012three,tullius2015superchiral,schaferling2012tailoring}. However, these fabrication processes are often complex, involving multiple steps under tightly controlled conditions, therefore posing challenges to scalability and cost efficiency. 

Tuning the properties of optical devices after fabrication is a versatile alternative without the need for re-patterning. Examples include thermo-optic modulation \cite{lewi2019thermal,bosch2019polarization}, ion beam irradiation \cite{krasheninnikov2010ion,li2017ion}, chemical treatment \cite{kowalczyk2020functional,zhao2024post}, and electrically induced bias \cite{thyagarajan2016millivolt}. Among these techniques, thermo-optic effects (e.g., thermal annealing) provide accessible tuning via temperature-dependent refractive index change \cite{long2020enhanced, kuppadakkath2023precision}, offering low-cost and large area compatibility. However, the spatial selectivity is limited by thermal diffusion.  

Ion beam irradiation offers high reproducibility and precise depth control through careful selection on ion species, energy, fluence, and incident angle \cite{li2019review, wu2023recent}. When combined with a mask, 3-dimensional nanoscale localized modifications become achievable \cite{yun2000fabrication, brun2011rapid}. Ion irradiation induces structural defects through atomic displacements and cascade collisions, creating structural defects such as vacancies, interstitial defects, defect complexes, and amorphization \cite{stern1971band, zammit1994optical, giri2001crystalline, sundari2004optical}. The resulting defects generate localized states within the band gap  and broadened Urbach tail absorption \cite{sundari2004optical}. As defect density increases with fluence, optical absorption also increases in the respective wavelength range.  

Here, we demonstrate that circular dichroism of a chiral metasurface can be optimized by irradiation with a 200 keV Ne ion beam. By varying the ion fluence, we systematically tune the material damage, thereby gradually increasing the absorption. Polarization-resolved transmission measurements reveal the ion fluence dependent evolution of CD, quantitatively analyzed by retrieving the refractive index change via an optimization-based fitting algorithm. Further insights into chiroptical behavior are obtained through the analysis of the local field enhancement and multipolar decomposition confirming the critical role of loss modulation to the CD response.   

\section{Metasurface Design and Numerical Simulation of Absorption-Dependent Circular Dichroism}

\begin{figure}
\begin{center}
  \includegraphics[width=0.65\linewidth]{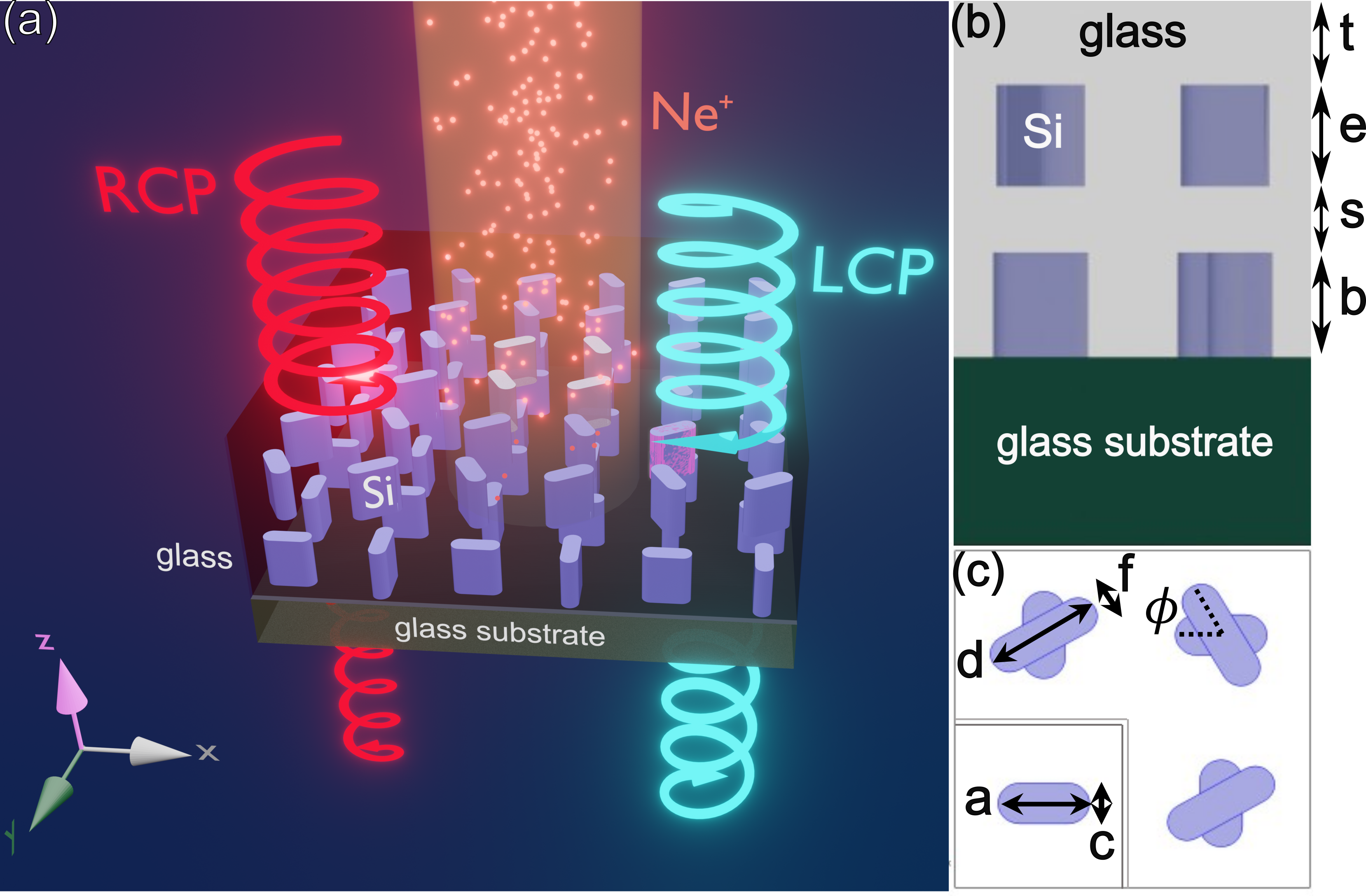}
  \caption{(a) Schematic illustration of the ion beam irradiated $C$\textsubscript{4}-symmetric chiral dielectric metasurface. Cross sectional images of the unit cell for the (b) side and (c) top view perspectives.}
  \label{fig:concept}
  \end{center}
  \end{figure}

The $C$\textsubscript{4} symmetric chiral metasurface under study is shown in Figure \ref{fig:concept}(a). This structure offers strong CD due to its three-dimensional broken structural symmetry while exhibiting circular eigenpolarizations owing to its 4-fold rotational symmetry \cite{tanaka2020chiral}. The metasurface consist of periodic arrays of twisted dielectric dimers. Each dimer is composed of 2 silicon nanocuboids arranged in an upper and a lower layer. The dimers are arranged on a two-dimensional square lattice in the metasurface plane (xy-plane) with lattice constant $p$. The upper and lower nanocuboids are separated by a distance of $s$ along the z-axis and oriented with a relative azimuthal angle of $60^\circ$ between the projection of their major axes on the metasurface plane. Adjacent dimers are rotated by $90^\circ$ with respect to each other. The cuboids consist of amorphous silicon and are situated on a silica glass substrate and embedded in a glass matrix. Figures \ref{fig:concept}(b) and (c) show a side and a top view, respectively, of the unit cell of the chiral metasurface. The lower and upper cuboids have dimensions of $a \times b \times c $ and $d \times e \times f$, respectively, and the thickness of the glass layer on top is $t$. The maximum of CD supported by this structure originates from the excitation of a chiral supermode extending over an entire super cell of the 4-fold symmetric structure \cite{tanaka2020chiral}.

Loss plays a crucial role as the inherent connection between loss and CD quantitatively determines the maximum achievable dichroism \cite{gorkunov2020metasurfaces}. Importantly, even if only aiming at different transmissions for RCP and LCP light, finite losses are vital to break time-reversal symmetry. Without these losses, the four-fold rotational symmetry of the structure ensures identical reflections for RCP and LCP light, preventing any transmission contrast \cite{kaschke2015helical}. The circular eigen polarization states and their coupling to the incident and outgoing waves can be analytically described in coupled-mode theory (CMT) under the slowly varying amplitudes approximation \cite{gorkunov2021bound,gorkunov2020metasurfaces,kondratov2016extreme}. For a metasurface with 4-fold-rotational symmetry, the transmission difference between the left- and right- circularly polarized light is explicitly given by \cite{gorkunov2020metasurfaces}:

\begin{equation}
\mathrm{CD}=\mathrm{T}\textsubscript{LCP\textrightarrow LCP} - \mathrm{T}\textsubscript{RCP\textrightarrow RCP} = 2\gamma_d \frac{\left|m\textsubscript{RCP}\right|^2-\left|m\textsubscript{LCP}\right|^2}{(\omega_0 - \omega)^2 + (\gamma_r+\gamma_d)^2.}
\label{eq: CMT_difference}
\end{equation}

Here, T\textsubscript{LCP\textrightarrow LCP} denotes the co-polarized transmission, defined as a transmission under LCP illumination and LCP detection, and, similarly, $T$\textsubscript{RCP\textrightarrow RCP} represents the transmission for RCP illumination and RCP detection. The parameters $m\textsubscript{LCP}$ and $m\textsubscript{RCP}$ are the coupling strengths of the corresponding polarizations to the considered resonant mode. In our chiral metasurface, the mode of primary interest is the chiral supermode, which exhibits strongly different coupling coefficients for RCP and LCP light due to its chiral nature. The total decay rate is given by $\gamma_r+\gamma_d$ where $\gamma_r$ is a radiative decay rate and $\gamma_d$ is a non-radiative or dissipative decay rate. In equation \ref{eq: CMT_difference}, $\omega$ denotes the frequency of the  incident light and $\omega_0$ is the resonance frequency of the chiral mode, such that CD reaches its maximum at $\omega_0$ under the condition that $\gamma_r$ and $\gamma_d$ are equal, known as critical coupling \cite{gorkunov2020metasurfaces}. 

To quantitatively investigate the role of the intrinsic loss via $\gamma_d$ on the CD response in the $C$\textsubscript{4}-symmetric chiral metasurface, we performed full-wave numerical calculations based on the finite element method (FEM) as implemented in the commercial software package COMSOL Multiphysics 6.2. Therein, we varied the imaginary part of refractive index of the silicon nanocuboids in the upper layer. Only the upper layer was modified to account for the finite ion penetration depth of 200 keV Ne irradiation (see SI Section B). The geometrical parameters were set as $a \times b \times c$ =  361 nm $\times$ 390 nm $\times$ 168 nm, $d \times e \times f$ = 464 nm $\times$ 335 nm $\times$ 153 nm, $s$ = 299 nm, $t$ = 46 nm and $p$ = 1400 nm. The complex-valued refractive index of silicon was modeled as $N_{s} = n_s + \textrm{i} \kappa_s$, where $n_s$ and $\kappa_s$ are the real and imaginary parts, respectively. Both the substrate and embedding glass were treated as lossless with refractive index $N_{g} = n_{g}$. To tune the absorption strength, we selectively modified the imaginary part of the refractive index of silicon in the upper layer expressed as $N_s^\mathrm{u} = n_s + \textrm{i}\alpha_s^\mathrm{u} \kappa_s$, where the superscript $\mathrm{u}$ denotes the upper layer and $\alpha^\mathrm{u}_s$ is the scaling factor for absorption. All refractive index values for silicon $n_s$, $\kappa_s$, and $n_g$ are provided in the Supplementary Information (SI) Figure S1.
\begin{figure}
\begin{center}
  \includegraphics[width=0.75\linewidth]{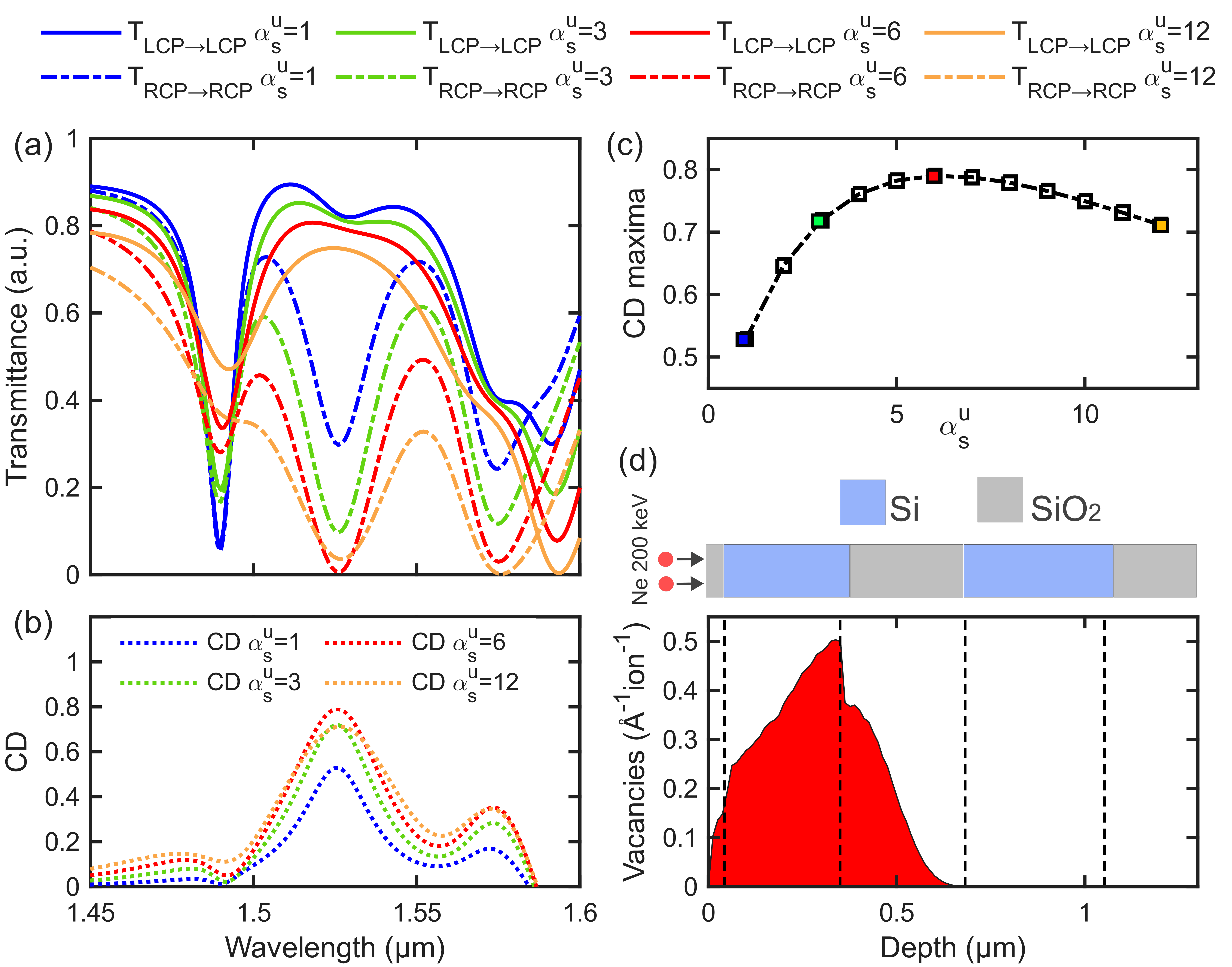}
  \caption{Calculated co-polarized transmission spectra of the $C$\textsubscript{4} chiral dielectric metasurface for different absorption scaling factor $\alpha^\mathrm{u}_s$ under (a) LCP (solid) and RCP (dashed) illumination. (b) Calculated circular dichroism spectra (c) CD peak as function of fluence, a maximum is shown at $\alpha^\mathrm{u}_s=6$. (d ) Vacancy profiles in the multilayer target simulated using a Stopping and Ranges of Ions in Matter (SRIM) simulation for 200 keV incident Ne.}
  \label{fig:idea_numerics}
\end{center}
\end{figure}

As shown in Figure \ref{fig:idea_numerics}(a), we observe a pronounced modulation of both LCP (solid) and RCP (dashed) transmission as $\alpha^\mathrm{u}_s$ increases. This, in turn, alters the CD spectra in Figure 2(b), calculated as the difference between transmittance of the two circular polarizations, namely $\mathrm{CD} = \mathrm{T}\textsubscript{LCP \textrightarrow LCP}-\mathrm{T}\textsubscript{RCP \textrightarrow RCP}$. Notably, the spectra exhibit three characteristic resonant features in the 1.45--1.60 \textmu m wavelength range with a prominent CD peak at 1.525 \textmu m. At $\alpha^\mathrm{u}_s = 6$, the RCP transmission approaches zero while the LCP magnitude remains nearly unchanged, resulting in a global CD maximum for our studied structure and  parameter range, indicative of critical coupling (Figure \ref{fig:idea_numerics}(c)). A further increase of $\alpha^\mathrm{u}_s$ reduces the CD again, consistent with an overdamped regime beyond critical coupling. Notably, however, this reduction happens very slowly, suggesting that the increase of intrinsic losses only mildly reduces the efficiency of chiral supermode excitation. For larger loss values we also note a significant broadening of the considered resonances as expected. 

\section{Evolution of Chiroptical Response Induced by Ion Irradiation}
The chiral metasurface (see SI Section F for fabrication details) was irradiated with 200 keV Ne ions at a room temperature. Neon was selected due to its chemical inertness to prevent any reaction with the target atoms. Ion induced damage was simulated using a Stopping and Ranges of Ions in Matter (SRIM) Monte Carlo\cite{ziegler1985stopping}. For simplicity, in these simulations the metasurface structure was approximated as a multilayer structure. To account for different possible ion trajectories in the actual metasurface and thus more closely reflect experimental conditions, two alternative multilayer configurations were investigated. Specifically, considering that the structure consists of upper and lower silicon cuboids tilted relative to each other, ions entering from the top glass can either (i) hit the upper silicon layer, or (ii) directly travel into the lower silicon layer. Figure \ref{fig:idea_numerics}(d) shows the vacancy distribution per incident ion over the target depth simulated for case (i). The layer thicknesses and materials of the model correspond to those in chiral metasurface geometry in Figure \ref{fig:concept}(b). The penetration depth of ions is $\sim$0.65 \textmu m with damage primarily concentrated in the upper three layers. The resulting vacancy distribution peaks inside the upper silicon layer at a depth of $\sim$0.38 \textmu m. Importantly, no damage is introduced in the lower silicon layer. The corresponding SRIM result for case (ii) is provided in SI Figure S3, showing that a small amount of vacancies can be created at the interface between glass and lower silicon layer.

We subsequently irradiated the chiral metasurface at six different fluences ranging from $10^{13}$ to $10^{15}$ ions cm\textsuperscript{-2}. Specifically, the ion fluences delivered in successive irradiation cycles were 0 (F0, pristine, before irradiation), $5\times10\textsuperscript{13}$ (F1), $10\textsuperscript{14}$ (F2), $2\times10\textsuperscript{14}$ (F3), $4\times10\textsuperscript{14}$ (F4), $7\times10\textsuperscript{14}$ (F5), and $10\textsuperscript{15}$ (F6) ions cm\textsuperscript{-2}, where the label F denotes the acummulated Ne fluence. After each ion irradiation cycle, polarization-resolved transmittance measurements were performed to evaluate the influence of the ion beam induced damage on the chiroptical properties of the metasurface (see SI, Section C).

 \begin{figure}
 \begin{center}
  \includegraphics[width=0.75\linewidth]{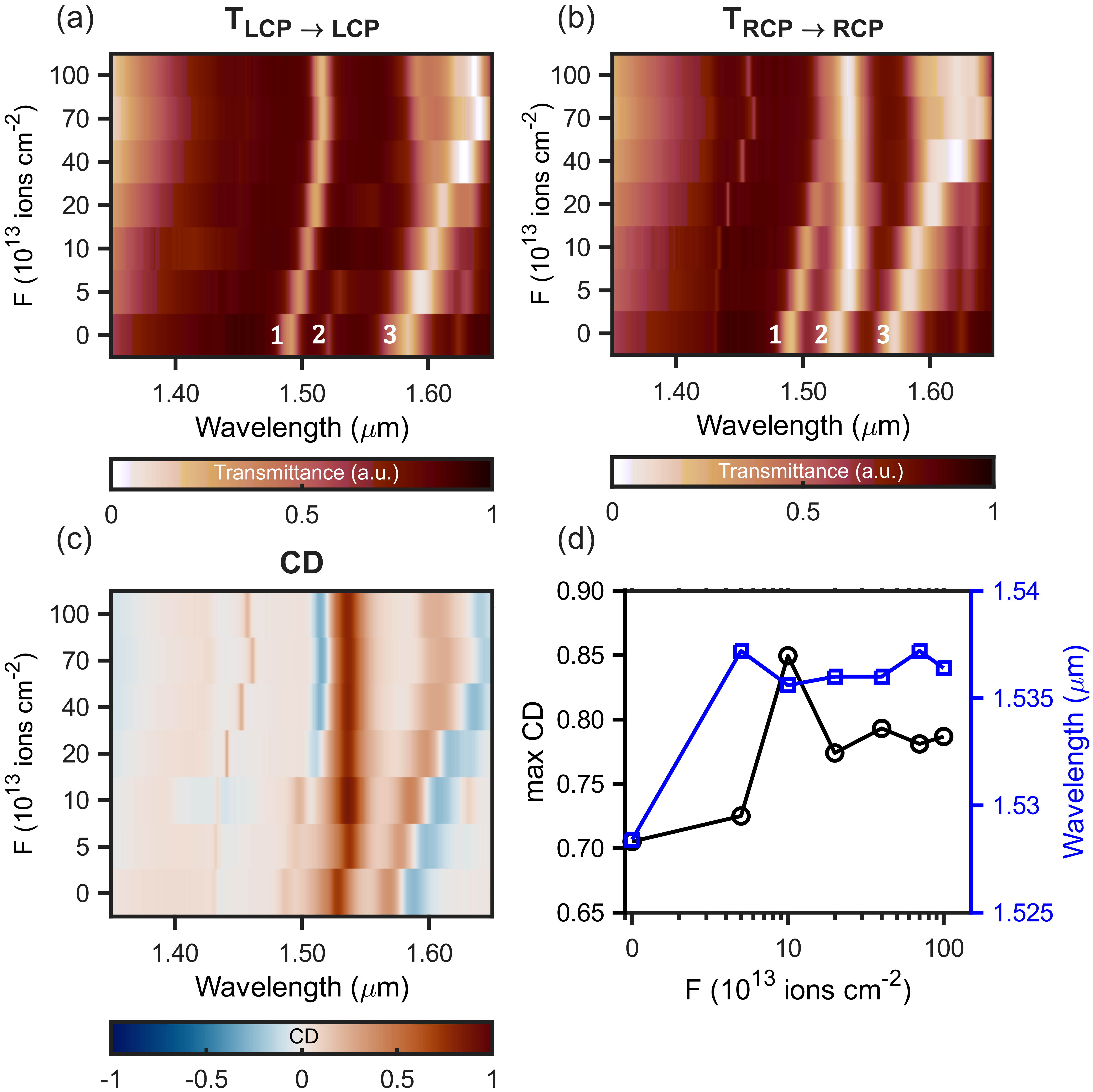}
  \caption{Measured polarization-resolved transmittance and circular dichroism for the $C$\textsubscript{4}-symmetric chiral metasurface under normal incidence illumination at varying ion fluences. (a) Co-polarized transmittance for LCP light, (b) co-polarized transmittance for RCP light. The resonance positions are marked with labels 1, 2, and 3. (c) CD spectra calculated as $\mathrm{T}_{LCP\rightarrow LCP}$-$\mathrm{T}_{RCP\rightarrow RCP}$. (d) Evolution of CD peak magnitude and wavelength as functions of ion fluence.}
  \label{fig:observation}
  \end{center}
\end{figure}

Figure \ref{fig:observation}(a) and (b) show 2D color maps of the zeroth-order transmission under normally incident LCP and RCP illumination, respectively, as functions of wavelength and ion fluence. The spectra at zero fluence (F0) correspond to those of the pristine metasurface. Three distinct resonant features are marked as resonance 1, 2, and 3. Resonance 2 corresponds to the chiral supermode giving rise to the  global maximum in CD apparent in Figure \ref{fig:observation}(c). 

As the ion fluence increases, resonances 1 and 3 are redshifted, indicating their sensitivity to the change of material properties. In contrast, the spectral position of resonance 2, the chiral supermode, shows a notable red-shift only after the first irradiation (F1) moving from 1.527 to 1.538 \textmu m, while it remains unchanged after subsequent irradiations. The stability suggests that this chiral mode is more robust against ion beam irradiation as compared to the other resonances. In addition to the observed spectral shifts, the transmission amplitudes also vary with fluence, implying changes in the coupling efficiency between the incident electromagnetic fields and the circular polarization eigenstates. Notably, at 10\textsuperscript{14} ions cm\textsuperscript{-2} fluence (F2) the RCP wave is suppressed at 1.538 \textmu m while the LCP transmission remains high.

The evolution of the CD spectrum is demonstrated in Figure \ref{fig:observation}(c). At higher ion fluences, additional CD features emerge. Weak negative CD values appear in a narrow wavelength window of 1.51--1.52 \textmu m due to stronger LCP suppression at resonance 1. Other negative CD responses arise at longer wavelengths, where the LCP and RCP resonance dips are no longer spectrally overlapped (see Figures \ref{fig:observation}(a) and \ref{fig:observation}(b)). 

Figure \ref{fig:observation}(d) summarizes the evolution of CD peak magnitude (black) and the corresponding spectral position (blue) as functions of ion beam fluence. A maximum CD of 0.85 is achieved after irradiation with an ion fluence of 10\textsuperscript{14} ions cm\textsuperscript{-2} (F2), representing an absolute increase of 0.15 compared to that of the pristine chiral metasurface. Beyond this optimum, the CD value initially declines and then shows little variation. These observations indicate that among all investigated fluence values, at 10\textsuperscript{14} ions cm\textsuperscript{-2}, the chiral metasurface comes closest to critical coupling, where the RCP wave is selectively coupled to the chiral supermode of the metasurface. We also conclude that beyond critical coupling, the chiral supermode remains relatively robust to the ion beam induced optical loss. The system still retains polarization selectivity of absorption and the increased optical loss primarily leads to resonance broadening rather than a rapid reduction in CD.

\begin{figure}
\begin{center}
\includegraphics[width=1\linewidth]{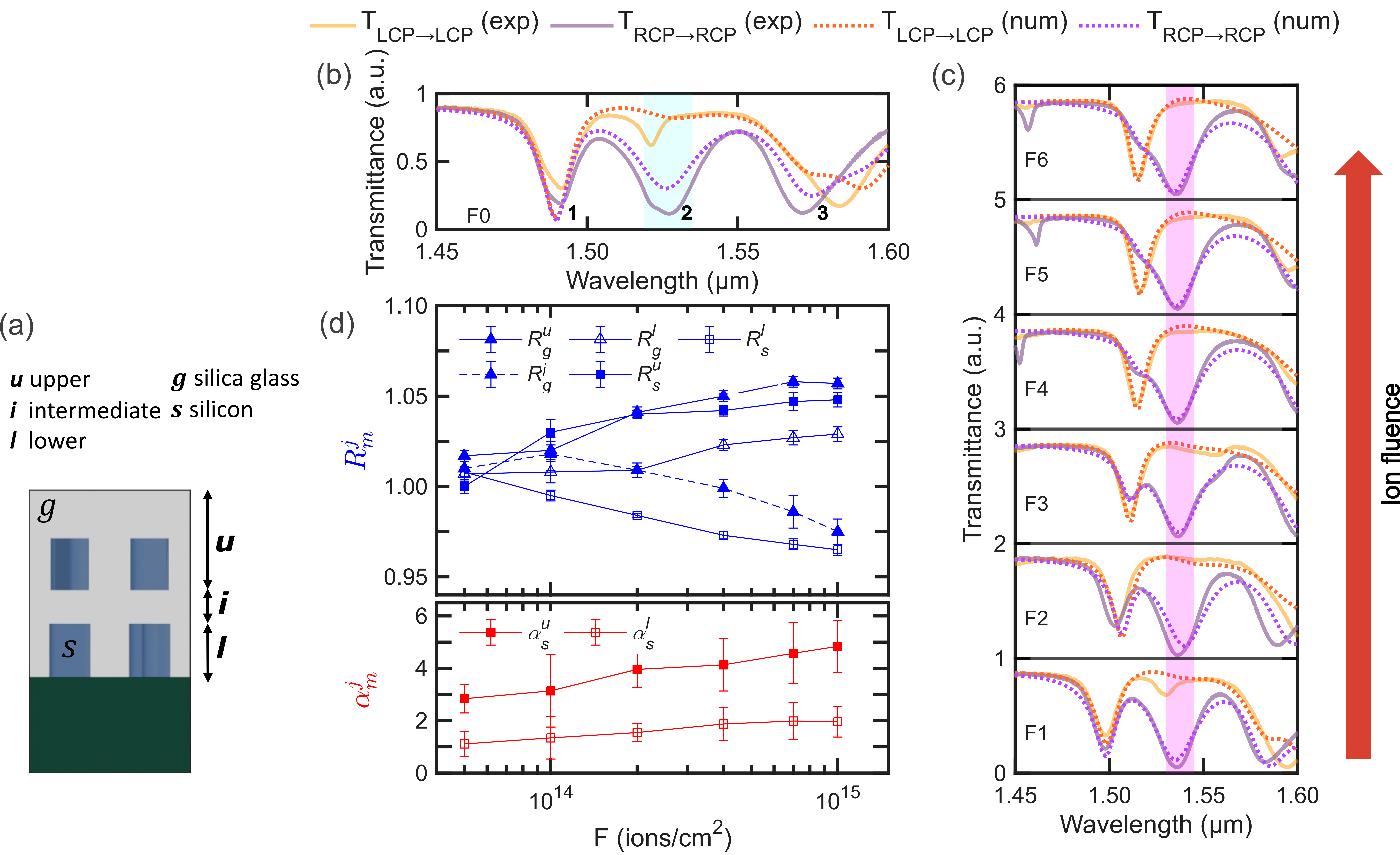}
\caption{ (a) Unit cell of the chiral metasurface used for retrieval of ion irradiation induced refractive index changes in the different layers and materials. (b) Experimentally measured (solid) and numerically calculated (dashed) results of the zeroth order transmission spectra for the pristine metasurfaces under LCP and RCP illuminations, with the three dominant features labeled as resonances 1, 2, and 3. (c) Experimentally measured (solid) and numerically calculated (dashed) zeroth order transmission spectra for the irradiated metasurface under LCP and RCP illuminations and for six different ion fluences (F1-F6). (d) Retrieved scaling factors for the real (blue) and imaginary (red) parts of the refractive indices for silicon and glass.}
\label{fig:numerics}
\end{center}
\end{figure}

To compare our experimental findings with theory and shed light on the changes in optical material properties underlying the observed CD evolution, we performed additional numerical calculations using the same tools and methods as described above. 
Figure \ref{fig:numerics}(b) compares the resulting calculated transmission spectra for RCP and LCP polarized light with corresponding experimental data for the pristine metasurface (F0), showing good agreement. Also, the retrieved values of $a$, $b$, $c$, $d$, $e$ and $s$ are fully consistent with scanning electron micrographs of the cross sections of the metasurface (see SI Table S1).  

For the irradiated metasurface, the refractive index changes induced by the ion beam were analyzed through a similar optimization framework. The optimization procedure focused on retrieving the refractive indices of the structural components of the metasurface able to reproduce the experimental results, while assuming the geometry of the metasurface to remain fixed. This retrieval procedure is complicated by the nonuniform, mainly depth-dependent, distribution of damage resulting from the ion irradiation, which leads to a depth-dependent variation of the refractive index within a given material. Thus, to capture the depth-dependent variations of the refractive index, while maintaining computational efficiency, the chiral metasurface was subdivided into three layers as illustrated in Figure \ref{fig:numerics} (a). The labels $j=\mathrm{u},i,l$ correspond to upper, intermediate and lower layer, respectively. Specifically, the layer $\mathrm{u}$ comprises the upper layer of silicon cuboids and the surrounding glass; the layer $i$ denotes the glass spacer layer separating upper and lower nanocuboids; and the layer $l$ consists of the lower layer of silicon nanocuboids and their embedding glass matrix. Next, within each layer the constituent materials are labeled as $m=s,g$ with $s$ for silicon and $g$ for glass. Using these labels, we can define a set of scaling factors quantifying the changes in complex refractive index in the different layers and materials as follows $R^j_m=\{R^\mathrm{u}_{g}, R^i_{g}, R^l_{g}, R^\mathrm{u}_{s}, R^l_{s} \}$ and $\alpha^j_m=\{\alpha^\mathrm{u}_{s}, \alpha^l_{s}\}$. For example, the parameter $R^\mathrm{u}_s$ describes the change of the real part of silicon refractive index in the upper nanocuboid while $\alpha^\mathrm{u}_s$ corresponds to its imaginary part. Optical loss was assumed to only occur in the silicon as the absorption of glass is expected to increase predominantly in the ultraviolet range \cite{zhu1998defects, leon2009neutron, martin2014optical}. To further validate this assumption, 1-mm-thick glass substrates were characterized after ion irradiation at multiple fluences. All samples exhibited nearly identical transmission spectra within 1.45--1.60 \textmu m wavelength range, confirming that the primary source of absorption in the near-infrared arises from the ion beam induced damage in silicon. 

In total, the expression of the modified refractive index for the silicon cuboids is given by $N^{j=l,\mathrm{u}}_{s} = R^{j=l,\mathrm{u}}_s n_s + \textrm{i} \alpha^{j=l,\mathrm{u}}_s \kappa_s$, where $n_s$ and $\kappa_s$ are the real and imaginary parts of the refractive index in pristine silicon. Similarly, the refractive index of glass is modeled as $N^{j=l,i,\mathrm{u}}_{g} = R^{j=l,i,\mathrm{u}}_g n_g$ where $n_g$ is the refractive index of pristine glass. To improve computational efficiency, the optimization was restricted to the wavelength range of 1.45 to 1.60 \textmu m. Within this range, the parameters $R^j_m$ and $\alpha^j_m$ are wavelength-independent and spatially homogeneous for each material inside a given layer. 

The comparison between the resulting numerically calculated (dashed line) and experimentally measured transmission spectra for varying ion fluence (F1-F6) are displayed in Fig. \ref{fig:numerics}(c), showing a good overall agreement. The notable improvement in the agreement as compared to Fig. \ref{fig:numerics}(a) (pristine sample) can be explained by the details of the fitting procedure. Since for the irradiated samples we allow for a variation of complex refractive indices, the procedure effectively takes not only changes in the actual material properties but also the influence of unavoidable scattering losses from interface roughness and other fabrication inaccuracies into account. For the pristine sample, in contrast, the optical material properties were kept fixed at experimentally measured values for a thin film. Despite the improved agreement, noticeable discrepancies appear in the longer wavelength range, likely arising from the assumption of nondispersive scaling factors, which, e.g., neglect the dispersive absorption behaviour as associated with sub-band-gap states in silicon \cite{zhu1998defect,oliviero2006damage}.

Moreover, it is important to note that the model used in the optimization process does not account for any possible geometrical changes induced by ion irradiation, including modification of nanostructure dimensions and interface morphology. In reality, the high fluence exposure may lead to structural changes in the target, such as deformation or shrinkage of nanostructures \cite{awazu2008elongation,klimmer2009size}, increased surface roughness, void formation, and interfacial mixing \cite{heinig2003interfaces, djurabekova2020defect}. Additionally, the retrieved parameters $R^j_{m}$ and $\alpha^j_{m}$ were treated as effective values, describing uniform distribution within each material of a given layer, thereby neglecting probable spatial variations in the refractive index distribution. Altogether, these simplifications may result in inaccuracies in the retrieved refractive index scaling factors. Nevertheless, apart from  overall good agreement achieved between the numerically calculated transmission spectra and corresponding experimentally measured data as already discussed (see Figure \ref{fig:numerics}(c)), the retrieved scaling factors also exhibit consistent trends as a function of fluence. 

Figure \ref{fig:numerics}(d) summarizes the fluence-dependent retrieved refractive index change factors $R^j_{m}$ and $\alpha^j_m$. The real components in the upper layer for both silicon and glass, $R^\mathrm{u}_s$ and $R^\mathrm{u}_g$, increase with increasing ion fluence, which is in accordance with the observed redshift of the modes (see Fig. \ref{fig:numerics}(d)) and can be attributed to ion induced densification of the materials. Such a densification result from compaction of the amorphous network due to the induced internal stress \cite{douillard1996swift,zheng2006densification}, atomic rearrangement \cite{an2006vacancy}, and bond distortion or breakage at molecular level \cite{devine1993ion,shojaee2017ion}. At the highest fluence (F6, 10\textsuperscript{15} ions cm\textsuperscript{-2}) the parameters for $R^\mathrm{u}_{s}$ and $R^\mathrm{u}_{g}$ were increased by approximately $4.8\pm0.4\%$ and $5.7\pm0.3\%$, respectively. Simultaneously, we observe a rapid increase in the absorption scaling factor $\alpha^\mathrm{u}_s$, implying substantial damage formation in the upper silicon cuboids consistent with the SRIM simulation shown in Figure \ref{fig:idea_numerics}(d). At the fluence of 10\textsuperscript{15} ions cm\textsuperscript{-2}, the imaginary part of silicon refractive index was strongly increased by a factor of $4.8\pm1.0$

\begin{figure}
  \includegraphics[width=1\linewidth]{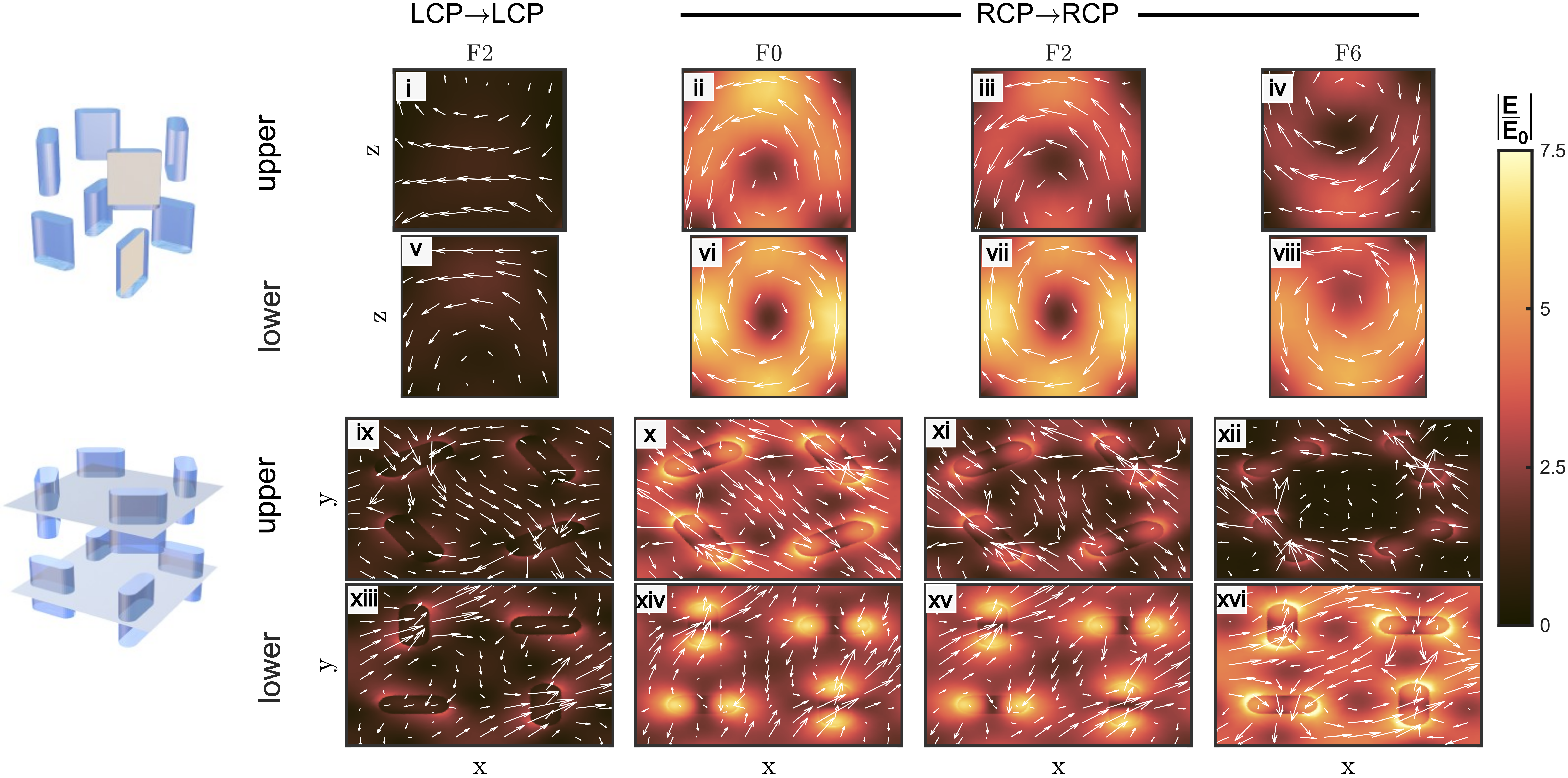}
  \caption{Normalized field intensity $\left|E/E_0\right|$ calculated in the upper and lower cuboids shown on out-of-plane (top) and in-plane (bottom) cross sections. Panels i,v,ix,xiii show $\left|E/E_0\right|$ under LCP excitation calculated for optimized parameters at F2 fluence and $\lambda$ = 1.538 \textmu m. Panels ii--iv, vi--viii, x--xii, and xiv--xvi show $\left|E/E_0\right|$ under RCP excitation evaluated in the pristine metasurface (F0) at 1.527 \textmu m and in the irradiated metasurface (F2 and F6) at 1.538 \textmu m, respectively.}
  \label{fig:fields}
\end{figure}

Figure \ref{fig:fields} (i,v,ix,xiii) and (iii,vii,xi,xv) show the normalized electric field intensity, $\left|E/E_0\right|$, in both out-of-plane and in-plane cross sections for the upper and lower cuboids at ion fluence F2 and $\lambda$=1.538 \textmu m, for LCP and RCP light, respectively. The color maps represent the field enhancement magnitude on the cross sections and the overlaid white arrows denote the tangential components of the electric field vectors. The calculation was evaluated on the cross sections parallel to z-direction (out-of-plane) at panels i,iii,v,vii and parallel to the metasurface plane (in-plane) at panels ix,xi,xiii,xv. In accordance with prior work \cite{tanaka2020chiral}, distinct circulating electric field profiles are observed at panels i,iii,v,vii, indicating the excitation of magnetic dipole modes within the individual silicon cuboids. Moreover, we record a strong contrast in the field enhancement between LCP and RCP illumination, consistent with preferential coupling of the absorptive chiral structures to the RCP wave giving rise to CD. Also, for RCP illumination, the near fields are strongly confined within the nanocuboids. Note that the chiral supermodes extends over the entire unit cell, including both layers, making it more robust against changes in a subset of resonators, which may contribute to the smaller red-shift observed for this mode with increasing ion fluence as compared to other modes.


Next, Figure \ref{fig:fields}(b) panels ii--iv, vi-viii, x--xii, xiv--xvi compare $\left|E/E_0\right|$ for different ion fluence values under RCP excitation at the corresponding chiral resonance. The calculations were performed at 1.525 \textmu m for F0 and at 1.538 \textmu m for fluences F2 and F6. Notably, the strong electric field circulation is preserved with increasing fluence (ii--iv and vi--viii). However, the field enhancement in the upper cuboid (ii--iv and x--xii) gradually decreases due to increased intrinsic loss induced by the ion irradiation. At the highest ion fluence (F6), the field localization in the lower cuboid slightly declines compared to that of F2 (xv and xvi). Altogether, these observations are in line with a reduction of the efficiency of RCP coupling and a decrease of CD.

\section{Conclusion and Outlook}
This work explored the potential of ion beam irradiation as a post-fabrication strategy to maximize CD in chiral resonant dielectric metasurfaces via ion irradiation induced tuning  of the absorption losses of its constituent materials. The amount of absorption losses is of utmost importance for tailoring CD in such metasurfaces, as CD is expected to reach a maximum under critical coupling to the resonant chiral mode. Specifically, we irradiated $C_4$-symmetric chiral bilayer silicon metasurfaces exhibiting a chiral supermode with Ne ion beams at energies of 200\,keV and various ion fluences. Already in its pristine form, this metasurface provides strong CD of 0.70 at the chiral supermode. For the optimal ion fluence of 10\textsuperscript{14} Ne ions cm\textsuperscript{-2}, we were able to increase CD to 0.85. Under this condition, the RCP light is nearly fully absorbed while the LCP light is mostly transmitted. 

To shed light on the physical mechanisms causing the increase in CD, numerical optimization procedures were employed to retrieve the ion irradiation induced changes to the optical material parameters from the fluence-dependent transmission spectra. From this analysis, a significant increase for the imaginary part of refractive index in the upper silicon layer could be determined. This increase is consistent with SRIM simulation results. Moreover, the observation that the CD is maximized for a specific fluence value and decreases again slightly for even higher fluences is indicative of successful loss engineering around the critical coupling condition. 

Our finding establishes ion beam irradiation as a reliable tool for tailoring polarization-selective absorption in resonant dielectric chiral metasurfaces. The possibility to irradiate metasurfaces with focused ion beams providing a high spatial resolution of damage formation on the nm scale offers additional exciting opportunities for tailoring the response of metasurfaces post-fabrication, thereby offering important prospects far beyond chiral metasurfaces. Overall, this study underscores ion beam irradiation as a powerful method for engineering the optical responses of prefabricated metasurfaces.


\section{Numerical Simulation}
\threesubsection{Vacancy Distribution Simulation}\\
Stopping and Ranges of Ions in Matter (SRIM) Monte Carlo calculations were performed for Neon ions with energy of 200 keV and angle of incidence of 7$^{\circ}$. To this end, we employed the software package SRIM-2008 using the detailed calculation with Full Damage Cascades option, which is adopted to simulate the track of ions and the subsequent cascades based on the binary collision approximation Kinchin-Pease formalism \cite{kinchin1955displacement}. During the simulation, the ions move in a series of discrete steps and collide with the host atoms in a randomized fashion. The path of the ion illustrates  the statistical properties of the stopping process. The number of Neon ions involved in the calculation was 20,000, the random seed number was set to 71,683. The model for SRIM calculation is depicted in Figure \ref{fig:idea_numerics}(d) consisting of multilayer stack of glass and silicon with layer thicknesses of 43-307-332-370-200 nm for the sequence SiO\textsubscript{2}-Si-SiO\textsubscript{2}-Si-SiO\textsubscript{2} on SiO\textsubscript{2} substrate. Mass densities were set to 2.200 g cm\textsuperscript{-3} for SiO\textsubscript{2} and 2.285 g cm\textsuperscript{-3} for Si (see SI Section B for simulation details).

\threesubsection{Numerical Optimization of Metasurface Geometrical Parameters}\\
Numerical optimization aiming to optimize the agreement between experimental and numerical transmission spectra for a variation of metasurface geometrical parameters within experimentally reasonable bounds is conducted using a nonlinear least squares curve-fitting procedure implemented in MATLAB integrated with COMSOL Multiphyiscs 6.2. The metasurface transmittance acts as a model function. It is simulated in COMSOL with a given initial guess of geometrical parameters acting as model inputs. The objective function is then taken as the sum of the squares of the error, between the model function and the experimental data points, and is then iteratively minimized through a sequence of updated values of parametric inputs. The minimization process is implemented using Levenberg-Marquardt algorithm \cite{levenberg1944method, marquardt1963algorithm}. Convergence is defined by a step tolerance of 10\textsuperscript{-6} and a function tolerance of 10\textsuperscript{-10}, which provide a balance between computational efficiency and numerical accuracy (see SI Sections D and E). 

\section{Experimental Section}

\threesubsection{Ion Beam Irradiation}\\
Ne ion beam irradiation was performed at an ion energy of 200 keV at a room temperature and an incidence angle of 7$^{\circ}$. The metasurface sample has area of $4 \times 4$ mm\textsuperscript{2}, which was irradiated by six subsequent ion fluences ranging from 5$\times$10\textsuperscript{13} ions cm\textsuperscript{-2} to 10\textsuperscript{15} ions cm\textsuperscript{-2}.

\threesubsection{Transmission Spectroscopy}\\
The polarization-resolved transmission of the chiral metasurface was measured using a broadband halogen light source, with the incident polarization controlled by a linear polarizer and quarter-wave plate and the transmitted polarization analyzed using the same combination. The transmitted light was guided into a spectrometer to record both co- and cross-polarized signals. A detailed description of the optical setup and measurement procedure is provided in the SI Section C.

\medskip
\textbf{Supporting Information} \par 

\medskip
\textbf{Acknowledgements} \par 
We acknowledge useful discussions with D. Arslan and A. Barreda. We also thank M. Steinert, O. Rüger, and S. Hillmann for their support with the focused ion beam (FIB) characterization of the metasurface sample. This research was funded by the Deutsche Forschungsgemeinschaft (DFG), IRTG 2675 Meta-Active, 437527638. 

\medskip

\bibliographystyle{MSP}
\bibliography{advance}

\begin{thebibliography}{10}
\providecommand{\url}[1]{\texttt{#1}}
\providecommand{\urlprefix}{URL }

\bibitem{beychok1966circular}
S.~Beychok,
\newblock \emph{Science} \textbf{1966}, \emph{154}, 3754 1288.

\bibitem{ranjbar2009circular}
B.~Ranjbar, P.~Gill,
\newblock \emph{Chemical biology \& drug design} \textbf{2009}, \emph{74}, 2 101.

\bibitem{tedesco2015induced}
D.~Tedesco, C.~Bertucci,
\newblock \emph{Journal of pharmaceutical and biomedical analysis} \textbf{2015}, \emph{113} 34.

\bibitem{kiss2019simple}
E.~Kiss, V.~A. Szab{\'o}, P.~Horv{\'a}th,
\newblock \emph{Journal of Inclusion Phenomena and Macrocyclic Chemistry} \textbf{2019}, \emph{95} 223.

\bibitem{decker2007circular}
M.~Decker, M.~Klein, M.~Wegener, S.~Linden,
\newblock \emph{Optics letters} \textbf{2007}, \emph{32}, 7 856.

\bibitem{decker2010twisted}
M.~Decker, R.~Zhao, C.~Soukoulis, S.~Linden, M.~Wegener,
\newblock \emph{Optics letters} \textbf{2010}, \emph{35}, 10 1593.

\bibitem{ma2015planar}
X.~Ma, M.~Pu, X.~Li, C.~Huang, Y.~Wang, W.~Pan, B.~Zhao, J.~Cui, C.~Wang, Z.~Zhao, et~al.,
\newblock \emph{Scientific reports} \textbf{2015}, \emph{5}, 1 10365.

\bibitem{zhao2012twisted}
Y.~Zhao, M.~Belkin, A.~Al{\`u},
\newblock \emph{Nature communications} \textbf{2012}, \emph{3}, 1 870.

\bibitem{zheng2022compound}
H.~Zheng, M.~He, Y.~Zhou, I.~I. Kravchenko, J.~D. Caldwell, J.~G. Valentine,
\newblock \emph{ACS nano} \textbf{2022}, \emph{16}, 9 15100.

\bibitem{chen2018spin}
Y.~Chen, X.~Yang, J.~Gao,
\newblock \emph{Light: Science \& Applications} \textbf{2018}, \emph{7}, 1 84.

\bibitem{wang2020controlling}
X.~Wang, J.~Duan, W.~Chen, C.~Zhou, T.~Liu, S.~Xiao,
\newblock \emph{Physical Review B} \textbf{2020}, \emph{102}, 15 155432.

\bibitem{ali2023dielectric}
A.~Ali, H.~S. Khaliq, A.~Asad, J.~Akbar, M.~Zubair, M.~Q. Mehmood, Y.~Massoud,
\newblock \emph{RSC advances} \textbf{2023}, \emph{13}, 30 20958.

\bibitem{govorov2010theory}
A.~O. Govorov, Z.~Fan, P.~Hernandez, J.~M. Slocik, R.~R. Naik,
\newblock \emph{Nano letters} \textbf{2010}, \emph{10}, 4 1374.

\bibitem{andrews2020physical}
S.~S. Andrews, J.~Tretton,
\newblock \emph{Journal of Chemical Education} \textbf{2020}, \emph{97}, 12 4370.

\bibitem{nafie1976vibrational}
L.~A. Nafie, T.~Keiderling, P.~Stephens,
\newblock \emph{Journal of the American Chemical Society} \textbf{1976}, \emph{98}, 10 2715.

\bibitem{xu2023expanding}
C.~Xu, Z.~Ren, H.~Zhou, J.~Zhou, C.~P. Ho, N.~Wang, C.~Lee,
\newblock \emph{Light: Science \& Applications} \textbf{2023}, \emph{12}, 1 154.

\bibitem{gorkunov2020metasurfaces}
M.~V. Gorkunov, A.~A. Antonov, Y.~S. Kivshar,
\newblock \emph{Physical Review Letters} \textbf{2020}, \emph{125}, 9 093903.

\bibitem{hendry2010ultrasensitive}
E.~Hendry, T.~Carpy, J.~Johnston, M.~Popland, R.~Mikhaylovskiy, A.~Lapthorn, S.~Kelly, L.~Barron, N.~Gadegaard, M.~Kadodwala,
\newblock \emph{Nature nanotechnology} \textbf{2010}, \emph{5}, 11 783.

\bibitem{hentschel2012three}
M.~Hentschel, M.~Sch{\"a}ferling, T.~Weiss, N.~Liu, H.~Giessen,
\newblock \emph{Nano letters} \textbf{2012}, \emph{12}, 5 2542.

\bibitem{tullius2015superchiral}
R.~Tullius, A.~S. Karimullah, M.~Rodier, B.~Fitzpatrick, N.~Gadegaard, L.~D. Barron, V.~M. Rotello, G.~Cooke, A.~Lapthorn, M.~Kadodwala,
\newblock \emph{Journal of the American Chemical Society} \textbf{2015}, \emph{137}, 26 8380.

\bibitem{schaferling2012tailoring}
M.~Sch{\"a}ferling, D.~Dregely, M.~Hentschel, H.~Giessen,
\newblock \emph{Physical Review X} \textbf{2012}, \emph{2}, 3 031010.

\bibitem{lewi2019thermal}
T.~Lewi, N.~A. Butakov, J.~A. Schuller,
\newblock \emph{Nanophotonics} \textbf{2019}, \emph{8}, 2 331.

\bibitem{bosch2019polarization}
M.~Bosch, M.~Shcherbakov, Z.~Fan, G.~Shvets,
\newblock \emph{Journal of Applied Physics} \textbf{2019}, \emph{126}, 7.

\bibitem{krasheninnikov2010ion}
A.~Krasheninnikov, K.~Nordlund,
\newblock \emph{Journal of applied physics} \textbf{2010}, \emph{107}, 7.

\bibitem{li2017ion}
Z.~Li, F.~Chen,
\newblock \emph{Applied Physics Reviews} \textbf{2017}, \emph{4}, 1.

\bibitem{kowalczyk2020functional}
T.~Kowalczyk,
\newblock \emph{Polymers} \textbf{2020}, \emph{12}, 5 1087.

\bibitem{zhao2024post}
D.~Zhao, G.~Yu, M.~Ge, M.~Han, H.~Meng, W.~Xiong, L.~Wen,
\newblock \emph{Separation and Purification Technology} \textbf{2024}, \emph{346} 127470.

\bibitem{thyagarajan2016millivolt}
K.~Thyagarajan, R.~Sokhoyan, L.~Zornberg, H.~Atwater,
\newblock \emph{arXiv preprint arXiv:1607.03391} \textbf{2016}.

\bibitem{long2020enhanced}
L.~Long, S.~Taylor, L.~Wang,
\newblock \emph{ACS Photonics} \textbf{2020}, \emph{7}, 8 2219.

\bibitem{kuppadakkath2023precision}
A.~Kuppadakkath, {\'A}.~Barreda, L.~Ghazaryan, T.~Bucher, K.~Koshelev, T.~Pertsch, A.~Szeghalmi, D.~Choi, I.~Staude, F.~Eilenberger,
\newblock \emph{Nanomaterials} \textbf{2023}, \emph{13}, 11 1810.

\bibitem{li2019review}
W.~Li, X.~Zhan, X.~Song, S.~Si, R.~Chen, J.~Liu, Z.~Wang, J.~He, X.~Xiao,
\newblock \emph{Small} \textbf{2019}, \emph{15}, 31 1901820.

\bibitem{wu2023recent}
X.~Wu, X.~Luo, H.~Cheng, R.~Yang, X.~Chen,
\newblock \emph{Nanoscale} \textbf{2023}, \emph{15}, 20 8925.

\bibitem{yun2000fabrication}
W.~S. Yun, J.~Kim, K.-H. Park, J.~S. Ha, Y.-J. Ko, K.~Park, S.~K. Kim, Y.-J. Doh, H.-J. Lee, J.-P. Salvetat, et~al.,
\newblock \emph{Journal of Vacuum Science \& Technology A: Vacuum, Surfaces, and Films} \textbf{2000}, \emph{18}, 4 1329.

\bibitem{brun2011rapid}
S.~Brun, G.~Guibert, C.~Meunier, E.~Guibert, H.~Keppner, S.~Mikhailov,
\newblock \emph{Nuclear Instruments and Methods in Physics Research Section B: Beam Interactions with Materials and Atoms} \textbf{2011}, \emph{269}, 20 2422.

\bibitem{stern1971band}
F.~Stern,
\newblock \emph{Physical Review B} \textbf{1971}, \emph{3}, 8 2636.

\bibitem{zammit1994optical}
U.~Zammit, K.~Madhusoodanan, M.~Marinelli, F.~Scudieri, R.~Pizzoferrato, F.~Mercuri, E.~Wendler, W.~Wesch,
\newblock \emph{Physical Review B} \textbf{1994}, \emph{49}, 20 14322.

\bibitem{giri2001crystalline}
P.~Giri, S.~Tripurasundari, G.~Raghavan, B.~Panigrahi, P.~Magudapathy, K.~Nair, A.~Tyagi,
\newblock \emph{Journal of Applied Physics} \textbf{2001}, \emph{90}, 2 659.

\bibitem{sundari2004optical}
S.~T. Sundari,
\newblock \emph{Nuclear Instruments and Methods in Physics Research Section B: Beam Interactions with Materials and Atoms} \textbf{2004}, \emph{215}, 1-2 157.

\bibitem{tanaka2020chiral}
K.~Tanaka, D.~Arslan, S.~Fasold, M.~Steinert, J.~Sautter, M.~Falkner, T.~Pertsch, M.~Decker, I.~Staude,
\newblock \emph{ACS nano} \textbf{2020}, \emph{14}, 11 15926.

\bibitem{kaschke2015helical}
J.~Kaschke, L.~Blume, L.~Wu, M.~Thiel, K.~Bade, Z.~Yang, M.~Wegener,
\newblock \emph{Advanced Optical Materials} \textbf{2015}, \emph{3}, 10 1411.

\bibitem{gorkunov2021bound}
M.~V. Gorkunov, A.~A. Antonov, V.~R. Tuz, A.~S. Kupriianov, Y.~S. Kivshar,
\newblock \emph{Advanced Optical Materials} \textbf{2021}, \emph{9}, 19 2100797.

\bibitem{kondratov2016extreme}
A.~Kondratov, M.~Gorkunov, A.~Darinskii, R.~Gainutdinov, O.~Rogov, A.~Ezhov, V.~Artemov,
\newblock \emph{Physical Review B} \textbf{2016}, \emph{93}, 19 195418.

\bibitem{ziegler1985stopping}
J.~F. Ziegler, J.~P. Biersack,
\newblock In \emph{Treatise on heavy-ion science: volume 6: astrophysics, chemistry, and condensed matter}, 93--129. Springer, \textbf{1985}.

\bibitem{zhu1998defects}
Z.~Zhu, Y.~Jin, C.~Li, Y.~Sun, C.~Zhang, Q.~Meng,
\newblock \emph{Nuclear Instruments and Methods in Physics Research Section B: Beam Interactions with Materials and Atoms} \textbf{1998}, \emph{146}, 1-4 455.

\bibitem{leon2009neutron}
M.~Le{\'o}n, P.~Mart{\'\i}n, R.~Vila, J.~Molla, A.~Ibarra,
\newblock \emph{Fusion Engineering and Design} \textbf{2009}, \emph{84}, 7-11 1174.

\bibitem{martin2014optical}
P.~Martin, D.~Jimenez-Rey, R.~Vila, F.~S{\'a}nchez, R.~Saavedra,
\newblock \emph{Fusion Engineering and Design} \textbf{2014}, \emph{89}, 7-8 1679.

\bibitem{zhu1998defect}
Z.~Zhu, M.~Hou, Y.~Jin, C.~Liu, Y.~Wang, J.~Han,
\newblock \emph{Nuclear Instruments and Methods in Physics Research Section B: Beam Interactions with Materials and Atoms} \textbf{1998}, \emph{135}, 1-4 260.

\bibitem{oliviero2006damage}
E.~Oliviero, S.~Peripolli, L.~Amaral, P.~F.~P. Fichtner, M.~F. Beaufort, J.~F. Barbot, S.~E. Donnelly,
\newblock \emph{Journal of applied physics} \textbf{2006}, \emph{100}, 4.

\bibitem{awazu2008elongation}
K.~Awazu, X.~Wang, M.~Fujimaki, J.~Tominaga, H.~Aiba, Y.~Ohki, T.~Komatsubara,
\newblock \emph{Physical Review B—Condensed Matter and Materials Physics} \textbf{2008}, \emph{78}, 5 054102.

\bibitem{klimmer2009size}
A.~Klimmer, P.~Ziemann, J.~Biskupek, U.~Kaiser, M.~Flesch,
\newblock \emph{Physical Review B—Condensed Matter and Materials Physics} \textbf{2009}, \emph{79}, 15 155427.

\bibitem{heinig2003interfaces}
K.~Heinig, T.~M{\"u}ller, B.~Schmidt, M.~Strobel, W.~M{\"o}ller,
\newblock \emph{Applied physics A} \textbf{2003}, \emph{77} 17.

\bibitem{djurabekova2020defect}
F.~Djurabekova, C.~Fridlund, K.~Nordlund,
\newblock \emph{Physical Review Materials} \textbf{2020}, \emph{4}, 1 013601.

\bibitem{douillard1996swift}
L.~Douillard, J.~Duraud,
\newblock \emph{Nuclear Instruments and Methods in Physics Research Section B: Beam Interactions with Materials and Atoms} \textbf{1996}, \emph{107}, 1-4 212.

\bibitem{zheng2006densification}
L.~Zheng, Q.~An, R.~Fu, S.~Ni, S.-N. Luo,
\newblock \emph{The Journal of chemical physics} \textbf{2006}, \emph{125}, 15.

\bibitem{an2006vacancy}
Q.~An, L.~Zheng, S.-N. Luo,
\newblock \emph{Journal of non-crystalline solids} \textbf{2006}, \emph{352}, 30-31 3320.

\bibitem{devine1993ion}
R.~Devine,
\newblock \emph{Journal of non-crystalline solids} \textbf{1993}, \emph{152}, 1 50.

\bibitem{shojaee2017ion}
S.~Shojaee, Y.~Qi, Y.~Wang, A.~Mehner, D.~Lucca,
\newblock \emph{Scientific Reports} \textbf{2017}, \emph{7}, 1 40100.

\bibitem{kinchin1955displacement}
G.~Kinchin, R.~Pease,
\newblock \emph{Reports on progress in physics} \textbf{1955}, \emph{18}, 1 1.

\bibitem{levenberg1944method}
K.~Levenberg,
\newblock \emph{Quarterly of applied mathematics} \textbf{1944}, \emph{2}, 2 164.

\bibitem{marquardt1963algorithm}
D.~W. Marquardt,
\newblock \emph{Journal of the society for Industrial and Applied Mathematics} \textbf{1963}, \emph{11}, 2 431.

\end{thebibliography}

\end{document}